\documentclass[10pt]{iopart}  

\usepackage{orcidlink}

\expandafter\let\csname equation*\endcsname=\relax 
\expandafter\let\csname endequation*\endcsname=\relax 
\usepackage{amsmath}

\begin{document}
\title{Deep Bayesian Experimental Design for Quantum Many-Body Systems}

\author{Leopoldo Sarra$^{1,2}$\orcidlink{0000-0001-7504-8656}, and Florian Marquardt$^{1,2}$\orcidlink{0000-0003-4566-1753}}
\address{$^1$ Max Planck Institute for the Science of Light, Staudtstraße 2, 91058 Erlangen, Germany}
\address{$^2$ Department of Physics, Friedrich-Alexander Universität Erlangen-Nürnberg, Staudtstraße 5, 91058 Erlangen, Germany}
\ead{leopoldo.sarra@mpl.mpg.de}

\begin{abstract}
    Bayesian experimental design is a technique that allows to efficiently select measurements to characterize a physical system by maximizing the expected information gain.
    Recent developments in deep neural networks and normalizing flows allow for a more efficient approximation of the posterior and thus the extension of this technique to complex high-dimensional situations.
    In this paper, we show how this approach holds promise for adaptive measurement strategies to characterize present-day quantum technology platforms.
    In particular, we focus on arrays of coupled cavities and qubit arrays.
    Both represent model systems of high relevance for modern applications, like quantum simulations and computing, and both have been realized in platforms where measurement and control can be exploited to characterize and counteract unavoidable disorder. Thus, they represent ideal targets for applications of Bayesian experimental design.

    \vspace{2pc}
    \vspace{2pc}

\end{abstract}

\section{Introduction}

We are currently witnessing rapid scaling in the number of components for quantum technology platforms.
Fulfilling the promise of fault-tolerant quantum computation will eventually require millions of qubits, and while current implementations still fall far short of that goal, the specific challenges of scaling are already apparent for the present-day devices including on the order of hundred qubits~\cite{kim_evidence_2023}.
Other areas like integrated photonic circuits~\cite{wang_integrated_2020} are following a similar trajectory, for applications such as neuromorphic computing~\cite{wetzstein_inference_2020} or sensing and more fundamental studies in areas like topological transport~\cite{ningyuan_time-_2015}.
There, large networks of beamsplitters, waveguides and resonators, again with component counts on the order of dozens or hundreds, are being fabricated and deployed.
Similar large-scale networks have now been fabricated and investigated for coupled mechanical resonators, producing phononic circuits with local access to vibrational modes.
\ioptwocol

Characterizing any of these devices is a very important but nontrivial task, especially if it is to be done in an efficient manner, within a limited time budget~\cite{gebhart_learning_2023}.
Active learning~\cite{settles_active_2012}, in the form of optimal experimental design can help, provided that one can employ techniques able to deal with the large number of parameters that are to be determined.
These parameters comprise resonance frequencies of optical or microwave modes, couplings between components (e.g. between resonators and qubits or waveguides, or beamsplitter transparencies), nonlinearities, propagation phase shifts, matrix elements for the effect of external drives, and decay rates.
Many different measurement approaches can be drawn upon, each of them coming with its respective parameters that can be adjusted prior to each new measurement.
In linear devices, specific components of the scattering matrix can be measured by injecting waves in some port and performing a homodyne measurement on another port.
Here, the frequency would be a continuous parameter, while, depending on the setup, the choice of ports could be another, discrete parameter.
In nonlinear systems, such as circuits comprised of qubits, manipulation via pulses in Ramsey-type schemes is the natural choice, with a final projective measurement of one or several qubits.
The drive amplitude and pulse times would then represent the measurement parameters to be optimized.
Systems that couple qubits and resonators or waveguides can also be characterized via nonlinear transmission, with the amplitude and drive frequency being treated as adjustable.

In all these cases, there is unavoidable fundamental noise in the measurement outcomes, namely shot noise in the case of wave transmission measurements and quantum projection noise in the case of qubit measurements~\cite{clerk_introduction_2010}.
This noise can naturally be reduced by extending the measurement duration, increasing the wave amplitude, or repeating multiple times the qubit pulse sequence together with its final projective qubit measurement (multi-shot measurement).
As a consequence, the information gain per such extended measurement increases.
However, there are limits to the usefulness of this naive optimization: wave amplitudes can be increased only so far before entering nonlinear regimes or heating up the device.
In addition, one cannot keep measuring at only one single  choice of the measurement parameter, since that  will eventually only pinpoint a particular function of the many underlying setup parameters and not allow to resolve all the parameters individually.
That is where active learning, i.e. choosing next measurement settings efficiently, can offer true benefits.

In this paper, we investigate recent and promising techniques from machine learning~\cite{foster_variational_2019,kleinegesse_bayesian_2020} to efficiently and accurately make the best parameter prediction given past measurement, and propose the next best measurement to perform.
We show applications of these techniques to the above-mentioned quantum devices.
We focus especially on quantum systems, such as chains of coupled cavities and arrays of qubits.
The presented framework is nonetheless completely general and appliable in a similar way to other settings.

\begin{figure}
    \centering
    \includegraphics[width=0.9\linewidth]{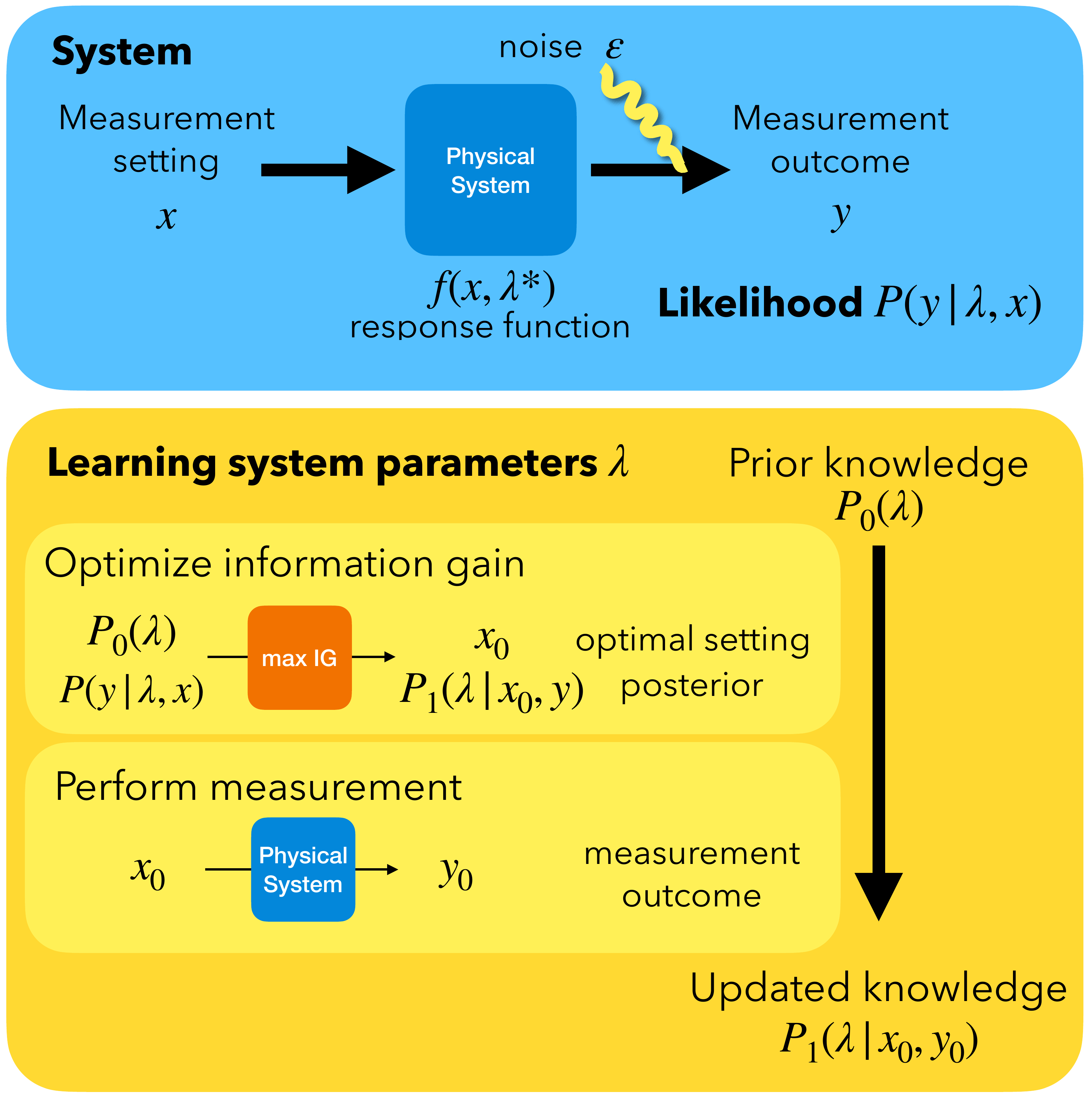}
    \caption{Sketch of the model of a system's response and summary of the parameter estimation procedure.
        At the top: the observed system produces a noisy observation given the measurement settings and the system's parameter $\lambda$.
        At the bottom: starting from the prior we have about the system, we can perform multiple steps.
        In each step, given the prior and the system's likelihood, we approximate the posterior distribution by minimizing the Barber-Agakov bound ~\eref{eq:information-gain0}.
        At the same time, the minimization also gives us the optimal measurement setting to perform the measurement at.
        After the measurement has been performed, the approximated posterior, conditioned on the measured value and outcome, becomes the new prior, and a new iteration can be performed.
    }
    \label{fig:algorithm-sketch}
\end{figure}

\section{Related Work}

The idea of finding the best experiment to perform is known as Active Learning in  the machine learning literature  \cite{cohn_improving_1994,settles_active_2012,tong_simon_active_2001} and as Bayesian optimal experimental design \cite{chaloner_bayesian_1995} in more specific parameter estimation applications \cite{fredlund_bayesian_2010, ryan_review_2016}.
It is very common in science to have a class of possible models, dependent on a set of parameters, and to perform experiments to find those that better describe the true system.
The outcome of each experiment does not give direct knowledge on the parameters themselves because of noise and measurement errors, but it provides partial information.
Therefore, it is generally not enough to perform as many measurements as the number of unknown parameters of the system, and more experiments are required.
When enough information is collected through measurements, we can identify the true parameters with a certain confidence.

When experiments are expensive, either in terms of cost, time and effort, it can be important to be efficient in the number of experiments required to characterize the given physical system.
In those cases, it is desirable to exploit the information obtained from each experiment as much as possible, and to choose the sequence of experiments in such a way that the smallest number of experiments is required.

Information Theory provides satisfactory answers to this problem, at least from the theoretical point of view.
In particular, the Bayes theorem shows the proper way to include a new measurement into our knowledge and update our predictions of the parameters of the system, and expected information gain~\cite{lindley_measure_1956} is the quantity to look at to find the best possible experiments to perform.

There have already been many remarkable applications in science~\cite{gal_deep_2017, sanchezlengeling_bayesian_2019} and in physics in particular, ranging from devising quantum metrological procedures \cite{hentschel_efficient_2011} to characterizing quantum dots \cite{lennon_efficiently_2019, nguyen_deep_2021}, superconducting quantum processors~\cite{hangleiter_precise_2021}, or sensors \cite{fiderer_neural-network_2021}, multi-phase estimation \cite{valeri_experimental_2020,cimini_deep_2023}, and their use is becoming more and more common.
Parameter estimation can then be further generalized to the complete design of the experiment. For example, ~\cite{krenn_automated_2016, melnikov_active_2018,krenn_computer-inspired_2020} completely automate the search for the experiment that solves a given task, in the case of quantum optics.

Nevertheless, optimal experimental design techniques have not been used in their full power up to recently, even though these techniques have been known for decades.
The main reason is that the exact evaluation of those statistical quantities can be very expensive.
Therefore, very rough approximations were usually made, including the use of empirical heuristics, the use of Gaussian processes \cite{liang_benchmarking_2021} and Gaussian posterior approximation \cite{vargas-hernandez_extrapolating_2018, duris_bayesian_2020} and the optimization through maximum likelihood estimation \cite{leclercq_bayesian_2018}.
The approximation usually depends on where the bottleneck is in practice, according to the specific problem: high-dimensional parameter space, large number of different possible experiments to perform, large number of measured quantities of each experiment, experiment execution time and cost, and total allowed number of experiments.

Modern neural networks \cite{goodfellow_deep_2016} that employ normalizing flows \cite{dinh_nice_2015, kobyzev_normalizing_2021} for the approximation of the posterior distribution, developed in the past few years, recently allowed for quite more efficient estimations \cite{foster_variational_2019,kleinegesse_bayesian_2020} that do not require such rough (and often unjustified) assumptions.
The price to pay for the increased precision of the approximation is clearly a larger computational effort.

\section{Methods}
\label{sec:methods}

Bayesian experimental design can be implemented as follows.
The settings of the experiment, which could be, for example, a measurement at a particular frequency, are described by a vector $x$.
The physical system is identified by hidden parameters $\lambda$, which we would like to determine.
Because of measurement noise, the outcome of the experiment, $y$, is not deterministic, but distributed according to a distribution $P(y|\lambda, x)$, called likelihood of an observation.
We can imagine, for example, $y = f(x,\lambda) + \epsilon$,  where $f$ is a deterministic function and $\epsilon$ is a random variable, e.g. Gaussian distributed.
We can update our knowledge of the parameters of the system with the Bayes rule~\cite{papoulis_probability_2009}:
given our inferred distribution $P_n(\lambda|\mathcal{M}_n)$ after $n$ measurements $\mathcal{M}_n=\{(x_1,y_1),\ldots, (x_n, y_n)\}$ (to which we will refer to as prior at step $n+1$), and the subsequent measurement $(x_{n+1}, y_{n+1})$, the updated distribution (posterior at step $n+1$) is
\begin{align}
    \label{eq:bayes}
    P_{n+1}(\lambda|\mathcal{M}_{n+1}) =\frac{P(y_{n+1}|\lambda, x_{n+1})P_n(\lambda| \mathcal{M}_n)}{P_n(y_{n+1}|x_{n+1})}.
\end{align}
The initial prior distribution $P_0(\lambda)$ is chosen arbitrarily, and it reflects our initial assumptions on the parameters of the system.

To choose the next measurement to perform, it is possible to define a query function, which assigns to each possible experiment $x$ its expected value: the $x$ for which the query function is maximized is the one that is expected to be the most useful to measure~\cite{settles_active_2012}.
A possible choice of the query function is the expected information gain when measuring at $x$:
\begin{align}
    \label{eq:information-gain}
     & I_n[\lambda,y](x)= \nonumber                                                                     \\
     & \;\;=\int dy d\lambda P_n(\lambda)P(y|\lambda,x) \log \frac{P_{n+1}(\lambda|y,x)}{P_n(\lambda)}.
\end{align}
This can be interpreted as the mutual information between $y$ and $\lambda$ given a measurement at $x$, or in other words, as the entropy reduction after measuring $(x,y)$.

However, Eq. \eref{eq:information-gain} is hard to estimate.
First, it requires to be able to sample from the prior distribution $P_0(\lambda)$, and second, it requires the value itself of the posterior probability distribution $P_1(\lambda|y_0,x_0)$, since it appears inside the logarithm.
In principle, we could obtain the posterior distribution with the Bayes rule, Eq. \eref{eq:bayes}, but it is not efficient to calculate its normalization factor $P_0(y|x)=\int d\lambda P_0(\lambda) \log P_0(\lambda)$.
Overall, to get Eq. \eref{eq:information-gain} we would require two nested Monte Carlo estimates, one for the evidence and one for the mutual information estimation.
Then, we should optimize over $x$ to get the best measurement to perform.

Modern neural network techniques allow implementing variational bounds and perform a much more efficient estimate~\cite{kleinegesse_gradient-based_2021}.
For example, we can avoid  evaluating the posterior~\cite{foster_unified_2020} explicitly by introducing a new function $Q(\lambda|y,x)$ and calculating the quantity
\begin{equation}
    \label{eq:information-gain0}
    I(x) = \int dy d\lambda
    P_0(y|x) P_1(\lambda|y,x) \log \frac{Q(\lambda|y,x)}{P_0(\lambda)}.
\end{equation}
If $Q$ is a probability distribution, we know that $KL(P_1(\lambda|y,x)||Q(\lambda|y,x))\geq 0$ thus
\begin{align}
    \label{eq:information-gain1}
    I(x) \leq I_0[\lambda,y](x).
\end{align}
This is the so-called Barber-Agakov bound of mutual information~\cite{barber_im_2003}.
We can parametrize $Q(\lambda|y,x)$ with a neural network, so that we do not need to find a different $Q$ function for every different $(x,y)$, but we can interpolate, amortizing the costs of the evaluation.

In particular, we use a conditional normalizing flow~\cite{winkler_learning_2019}: we start from a normal Gaussian distribution $\mathcal{G}(z)$ and perform a series of invertible transformations (parametrized by neural networks) having a Jacobian that is simple to estimate.
Optimizing over $Q(\lambda|y,x)$ means to optimize over the parameters of the neural networks that implement the transformations.
Normalizing flows can be easily implemented in common deep learning frameworks like TensorFlow \cite{martin_abadi_tensorflow_2015}.
In particular, we use TensorFlow Probability \cite{dillon_tensorflow_2017}.
The procedure is summarized in~\fref{fig:algorithm-sketch}.
Please refer to~\ref{app:bayes} for more details.

There are multiple ways to estimate the performance of the proposed result (and compare with alternative techniques).
The sum of the information gained at step $n+1$ after measuring $(x_{n+1},y_{n+1})$,
\begin{equation}
    \int d\lambda P_n(\lambda|y_n, x_n) \log\frac{Q(\lambda|y_n,x_n)}{P_n(\lambda)},
\end{equation}
can be a measure for how much we understood about the system.
However, in contrast to \eqref{eq:information-gain0}, which represents the expected information gain averaged over all the possible measurement outcomes, this quantity can also be negative. Indeed, it is possible that an unexpected outcome of a measurement increases the global uncertainty on the parameters.
Properties of the final posterior distribution such as its variance, or the comparison of its mean (or most likely value) with the true parameters are also a possible metric to consider.
On the other hand, the best prediction for an observation is actually the average over all the allowed parameters
\begin{equation}
    P(y|x) = \int d\lambda P_n(\lambda |\mathcal{M}_n) P(y|\lambda,x).
\end{equation}

It is possible to compare this prediction with the likelihood calculated at the true parameters $\lambda^*$, for example using the Kullback-Leibler divergence.
This gives an idea of how much the predictions of our model can differ from the observations.
It is important to emphasize that if the physical model is correct, we would expect only a single parameter set to be relevant, corresponding to the true parameters of the system.
However, if the model does not represent well the true system, the use of multiple values may be helpful to produce a better approximation.

\section{Applications and discussion}

We have already mentioned in the introduction the wide variety of quantum technology platforms, system parameters, and potential measurement approaches that could benefit in principle from active learning approaches.
In the following, we have chosen two illustrative examples that are sufficiently distinct in their characteristics.
As a first example, we consider a linear device, where wave transmission is measured to extract the resonance frequencies of a coupled-cavity array.
Such cavity arrays~\cite{hartmann_quantum_2008,houck_-chip_2012, ningyuan_time-_2015,altman_quantum_2021} have been considered in studies of transport as coupled-resonator optical waveguides in optical setups or quantum many-body physics of photons, when combined with nonlinear elements like qubits.
In our second example, we turn to an array of qubits, which is a widespread platform used for quantum simulations and quantum computing~\cite{houck_-chip_2012,bernien_probing_2017,zhang_observation_2017,ma_dissipatively_2019,altman_quantum_2021,zhang_superconducting_2023}.
There, we consider a typical pulsed scheme followed by a projective measurement to extract qubit parameters.
\ref{app:examples} provides additional technical details about the implementations of the examples.

\subsection{Coupled cavities}

\begin{figure}
    \centering
    \includegraphics[width=\linewidth]{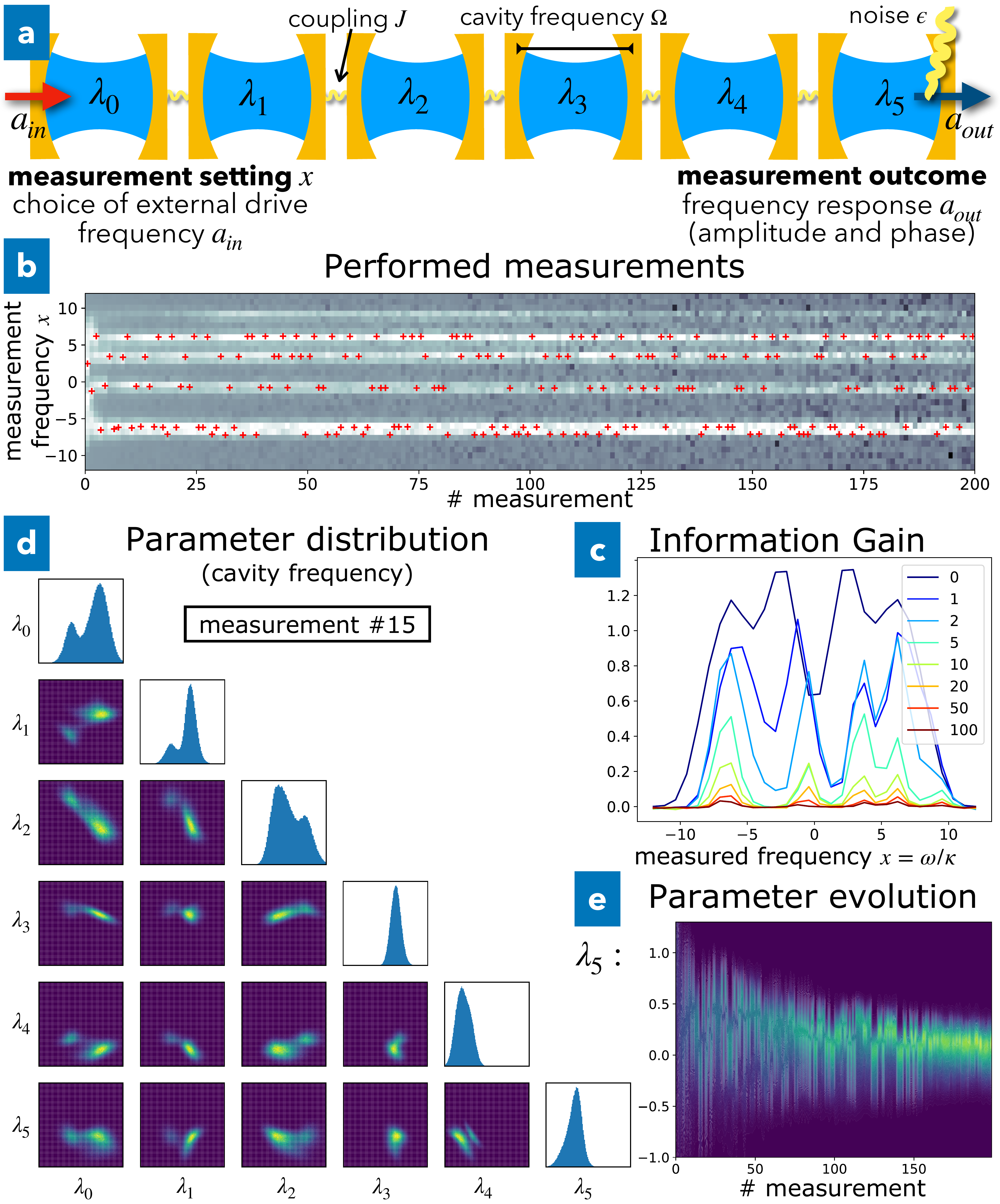}
    \caption{Deep Bayesian optimal experimental design applied to estimating the frequencies of an array of $6$ coupled cavities.
        (a) Sketch of the system.
        (b) Measurements at each step.
        The red symbols show the chosen frequency $x$ to measure at each step.
        In the background, the value of the expected information gain for each possible $x$ (brighter is higher), normalized at each step, i.e. $IG(x)/\max_x IG(x)$.
        (c) Expected information gain values for each possible measurement $x$. A peak represents an optimal value to measure. Different lines represent different measurement steps.
        (d) Inferred parameter distribution after $15$ measurements.
        The diagonal shows the marginal distribution $P(\lambda_i)$, the off-diagonal plots the two-variable slice $P(\lambda_i,\lambda_j)$.
        (e) Evolution of the marginalized posterior distribution of the last cavity frequency, until converging to a sharp peak around the true value.
    }
    \label{fig:measurements}
\end{figure}

We consider a cavity array with the following Hamiltonian:

\begin{equation}
    {\hat H}=\sum_j \Omega_j {\hat a}_j^{\dagger} {\hat a}_j +  \sum_j J_{j+1}({\hat a}_j^{\dagger} {\hat a}_{j+1} + {\rm h.c.}),
\end{equation}
where the sum runs over all but the last of the cavity modes in this chain with open boundary conditions.
Here and in what follows, we consider $\hbar =1$.

Since the setup we are dealing with is linear, it is sufficient to solve the classical equations of motion for the coupled modes, driven by a wave entering from a waveguide coupled to the first cavity.
To this end, we consider the classical coherent-state amplitudes $a_j$ corresponding to the quantum operators ${\hat a}_j$, including drive and decay as prescribed by input-output theory~ \cite{measurement}.
For brevity of our notation, we collect all these amplitudes in a vector and all frequencies in a matrix, where first-neighbour interactions are $J$ and proper frequencies $\Omega_i$.
Let our system also have some internal decay $\kappa_\text{int}$ and external decay  $\kappa_\text{ext}$, which only applies to the cavities coupled to the environment, for example the first and last.

Given the entering fluctuating field $a_\text{in}$, which can also contain a laser drive, the output field $a_\text{out}$ is given by~\cite{cohadon_quantum_2020}
\begin{equation}
    \label{eq:response}
    a_\text{out} = \mathcal S a_\text{in} =  \left( \mathbf 1 - \frac{\sqrt{\kappa_\text{ext}}}{-i(\omega - \Omega) + \frac{\kappa}{2}}  \right) a_\text{in}.
\end{equation}

In our example, we imagine driving the first cavity and looking at the response at the last cavity.
Our goal is to extract the scattering matrix element for transmission from the first to the last cavity, given by the $\mathcal S_{0N}$ in ~\eqref{eq:response}.
Being it a complex number, its knowledge corresponds to measuring both the amplitude and phase of the emitted light.
By performing the smallest possible number of measurements at a frequency $\omega$ close to the resonance frequency, we want to discover the frequencies of all the cavities.
For example, they may all be close to a common resonance frequency, but slightly detuned.
Each measurement is affected by noise $\epsilon$, which can be assumed to be Gaussian \cite{clerk_introduction_2010}.
This noise, being observed in a measurement of the field amplitude, ultimately arises from quantum vacuum noise being injected into the input ports of the device (and through the dissipation channels) and propagating linearly through the setup.

In~\fref{fig:measurements}, we show our results.
We consider $6$ coupled cavities, with each random coupling uniformly chosen in $J_j/\kappa\in[1,3]$.
We assume a setting where the couplings $J$ and the dissipation $\kappa$ have already been calibrated and are therefore known, and our goal is to find the detunings.
In the general notation of section~\ref{sec:methods}, the measurement setting $x$ corresponds in this case to the choice of the drive frequency $\omega$, the response function $f(x,\lambda)$ corresponds to the scattering matrix element $\mathcal{S}_{0N}$, and the unknown parameters of the system $\lambda$ are the frequencies of the cavities $\Omega_i$.

As more measurements are performed, we improve our estimate of the cavity frequencies.
We see that the information provided by new measurements decreases as we make more observations.
At the same time, the posterior distribution converges to a sharp peak around the true values. For example, \fref{fig:measurements}e shows the evolution of the last cavity frequency distribution.
It is important to emphasize that even if the posterior eventually converges to a Gaussian, it is not Gaussian at the beginning, as shown in ~\fref{fig:measurements}d.
This is why it is necessary to approximate the posterior using a neural network approximation.

By looking at the measurement frequencies that the active approach selects, we can identify a pattern, or measurement strategy: it measures the frequencies where the slope of the response is larger, alternating between different points.
These points give indeed the largest information on the intrinsic frequencies of the cavities.

\begin{figure}
    \centering
    \includegraphics[width=0.95\linewidth]{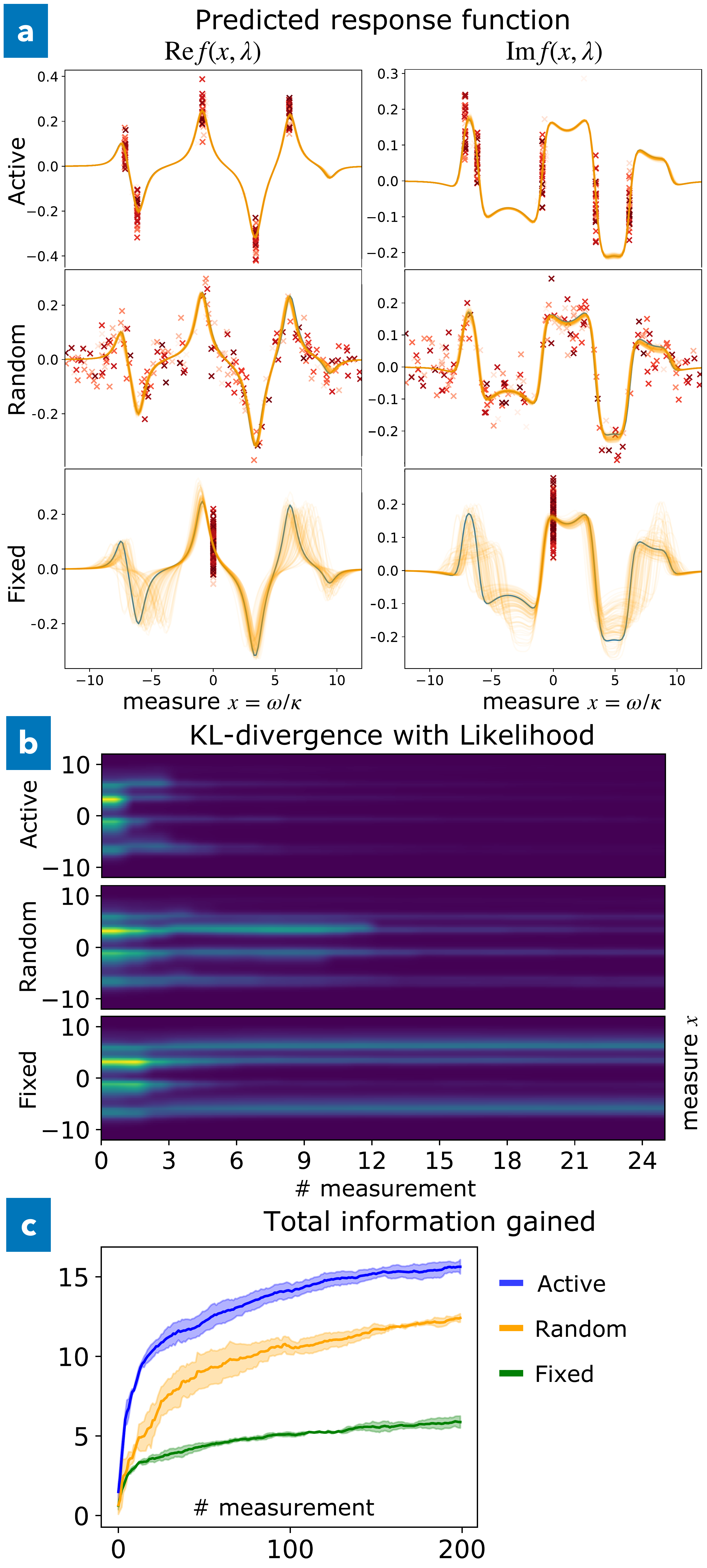}
    \caption{Comparison of different measurement strategies for the coupled cavities system.
        (a) Final response functions after inferring the parameters with the active, random, or fixed strategy. The orange curves represent the response function induced by various samples of parameters $\lambda$ from the final posterior distribution. In blue, the true response. Red symbols mark the performed measurements and their outcome.
        (b) Evolution of the KL-divergence between the inferred response function distribution $P(y|x)$ and the system likelihood during the various measurements. Brighter means higher.
        (c) Cumulative sum of information gained after each measurement, i.e. $\sum_{n=1}^N  \log{P_n(\lambda|y,x)} - \log{P_{n-1}(\lambda)}$. Higher values represent learning more about the system. Results are averaged over $3$ runs.
    }
    \label{fig:strategies}
\end{figure}

We compare the results of the inference with an active choice of the measurement with two other strategies: fixed and random.
The fixed strategy is the simplest: all the measurements are performed at the same $x$ value, and it is clearly often not possible to learn all the parameters from that.
There will usually be a region of the response function, close to the measurement region, that is very well learned, while other regions will not be very accurate.
A better naive measurement selection approach is the random one.
In this case, measurements are chosen uniformly within the specified measurement range (here, $x=\omega/\kappa\in[-12,12]$).
We can see that in~\fref{fig:strategies}: the active strategy, i.e. the one that chooses the next measurement greedily by maximizing the expected information gain, learns the parameters of the system faster than the random one.
Asymptotically, both the random and active strategy will converge to the right values.
However, active selection allows to save a lot of measurements.


\subsection{Array of qubits}

We consider an array of $N$ qubits, whose many-body Hamiltonian reads like
\begin{equation}
    H=\sum_{i=0}^{N-1} \omega_i \sigma_z^{(i)} + J \sum_{i=0}^{N-2} \sigma_x^{(i)} \sigma_x^{(i+1)}.
\end{equation}
At the beginning, the system starts in its ground state $|\psi_0\rangle$.
We assume that qubits can be individually addressed, and can be subjected to microwave pulses and projective measurements.
In particular, we excite the first qubit of the array by applying a microwave pulse with pulse area $\pi$.
This corresponds to implementing the unitary
\begin{equation}
    U=e^{-i\frac{\pi}{2}\sigma_x^{(0)}}.
\end{equation}
We let the system evolve freely for some time $t$, and then we perform a projective measurement on the first qubit.
The evolution allows for the propagation of the external excitation.
The outcome of the measurement is binary, according to probability
\begin{equation}
    p_\uparrow^{(0)} =\langle \psi(t) | \hat{\mathcal P}_{\uparrow}^{(0)} | \psi(t) \rangle,
\end{equation}
where $\hat{\mathcal P}_{\uparrow}^{(0)}= |\uparrow\rangle_0 \langle \uparrow|$ is the projection operator on the "up"-state of the first qubit (qubit number $0$).
In each experiment, we can choose the duration of the free evolution after the initial pulse, which we will denote as the "measurement pulse duration" $x\in[0,5]$.
By performing the smallest possible number of experiments, we want to discover the frequencies $\omega_i$ of all the qubits.

\begin{figure}
    \centering
    \includegraphics[width=\linewidth]{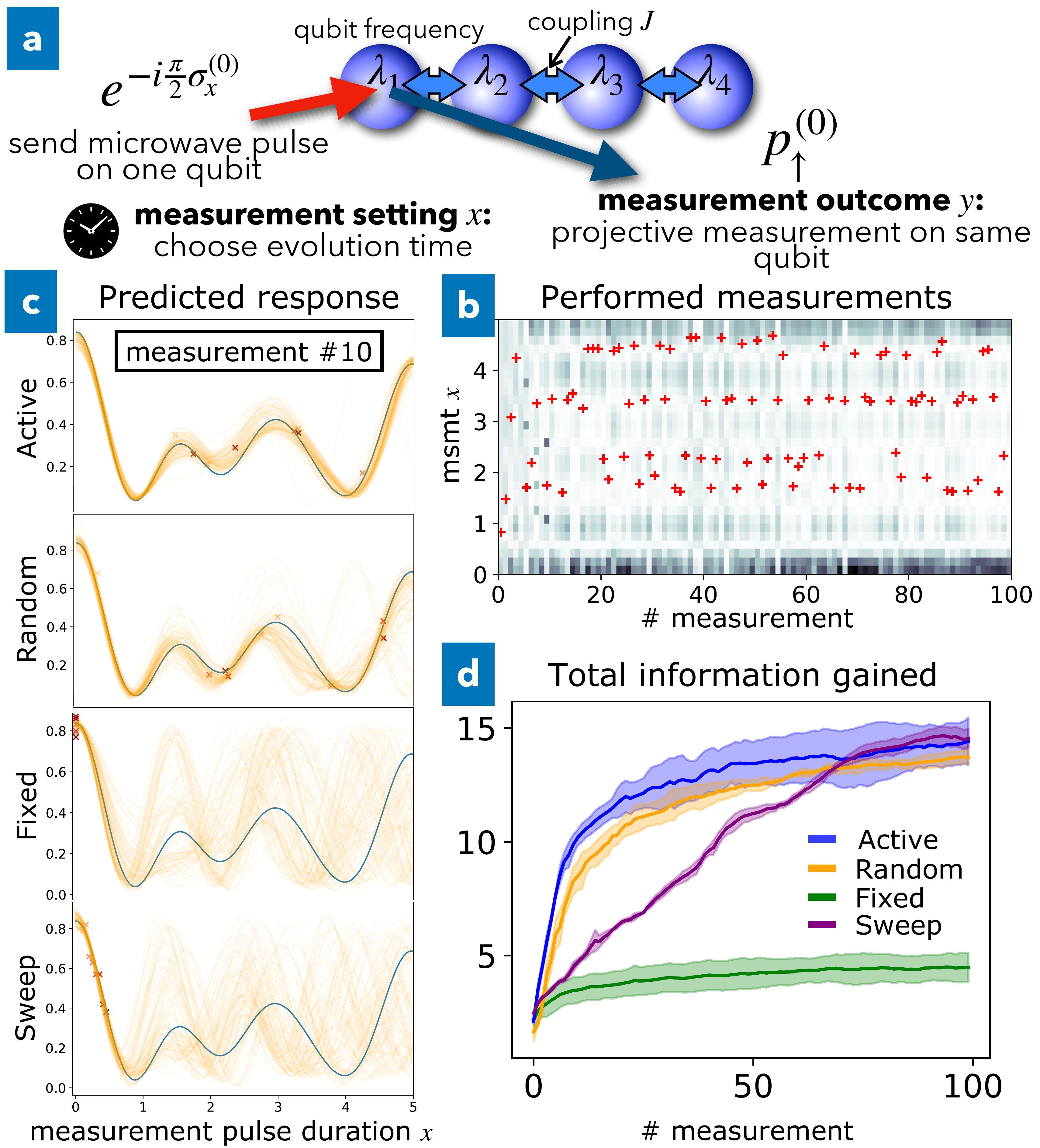}
    \caption{Deep Bayesian experimental design applied to the estimation of the frequencies of an array of $4$ qubits.
        (a) Sketch of the system.
        (b) Chosen pulse duration $x$ at each step.
        The red symbols show the chosen $x$ to measure at each step. In the background, the value of the expected information gain for each possible $x$ (brighter is higher), normalized at  each step maximum.
        (c) Response functions after $10$ measurements, comparing the active, random, fixed and uniformly spaced approaches. The orange curves represent the response function induced by various samples of parameters $\lambda$ from the final posterior distribution. In blue, the true response. Red symbols mark the performed measurements and their outcome.
        (d) Sum of information gained after each measurement, i.e. $\sum_{n=1}^N  \log{P_n(\lambda|y,x)} - \log{P_{n-1}(\lambda)}$.
        Higher values imply learning more about the system.
        Results are averaged over $3$ runs.
    }
    \label{fig:strategies-qubits}
\end{figure}

We considered a $4$-qubit system and performed $100$ measurements in series, choosing the pulse duration for each of them.
In this case, each of those measurements is in itself a multishot measurement, being repeated $100$ times to give better statistics (otherwise the outcome would only be either $0$ or $1$).
Again, as shown in ~\fref{fig:strategies-qubits}, we compare the active strategy with a random and a fixed one, and show the efficiency advantages of the active one in terms of gained information.
In this case, we also show the performance of a uniform strategy, which takes sequential equally-spaced measurements, i.e. sweeping the pulse duration from $0$ to the maximum allowed value.
It is interesting to observe that this strategy eventually learns the parameters of the system, but it requires many more measurements than the random one.
Also, in \fref{fig:strategies-qubits}d we see that the active strategy usually requires, on average, sensibly fewer measurements than the random strategy to provide the same total information gained. For example, to reach a total information gained of $10$, we require about $10$ measurements with the active approach or about $20$ measurements with the random one.

The fixed strategy always measures at $x=0$, i.e. does not allow any relaxation after the microwave pulse on the first qubit.
As a consequence, this measurement effectively only provides information about the interacting ground state of the system, which already contains some limited information about the qubit frequencies.
We emphasize, indeed, that since the interacting quantum many-body ground state is not an eigenstate of $\sigma_z$ acting on the first qubit, the probability of measuring $\uparrow$ at $t=0$ is not $1$.

Furthermore, it is possible to spot a sort of strategy in the pattern of the measurements performed by the active selection: a few measurement regions are alternated cyclically.
Indeed, we can imagine that some regions give the largest information about some parameters.
After we measure in one region, we increase the accuracy on some parameters, and another region becomes more relevant, until it is useful to measure again at the initial spot.
This cycle is repeated until we reach the desired accuracy on the parameters.

\section{Outlook}

In this paper, we have introduced deep optimal Bayesian experimental design for modern quantum technologies.
This approach approximates the posterior update with a variational bound and a deep neural network and allows extracting the optimal measurement to perform at each step.
We have shown the application of this technique to two promising quantum platforms, cavity arrays and qubit chains.
In both cases, the active measurement selection technique allows learning the parameters of the system with fewer measurements than other strategies.

Depending on the application, different assumptions should be made, leading to different tradeoffs between the approximation precision and the time efficiency.
The approach we propose is especially advantageous in the setting where each measurement is very expensive and it is worth spending a long calculation time to really find the optimal next measurement setting.

The main challenge at present is the time it takes to update the neural network representing the Bayes posterior distribution as well as optimizing over possible measurement settings.
This time is still too large in order for this technique to be deployed economically ``as is" in the given scenarios, where individual measurements can happen on microsecond time scales, and even extended sequences used to reduce shot noise will probably not last longer than a millisecond for one measurement setting, except when extreme accuracy is called for.
In the present (not yet very optimized) technique, an optimization run still takes on the order of hundreds of seconds per measurement step, which is currently a drawback of  Bayesian Optimal Experimental Design based on deep neural networks in general.
However, a remedy would consist in performing the optimization on a number of different example scenarios, in simulations, and then train a neural network in a supervised fashion to learn the suitable choice of the next measurement setting based on all the previous settings and outcomes.
In this way, we could really develop a sort of policy that can be deployed to characterize unknown systems of the same class.

A specific cost function may also be added to take into account that some measurement values may require more resources and thus be more expensive than others~\cite{settles_active_2012}.

A further generalization can be to drop the likelihood assumption and try to develop a likelihood-free active inference technique.
This would be similar to \cite{cimini_deep_2023}.
However, we do not want to discretize the input space $x$, since it would not scale up to large dimensions.
Finally, we notice that we are employing a greedy strategy which always chooses the single next best measurement to perform. Alternative approaches may include providing an overall measurement budget (e.g. total number of measurements) or set a target total information gain to achieve.
It would be interesting to study how these sequential strategies~\cite{foster_deep_2021}, perhaps approximated with a reinforcement learning agent~\cite{sutton_reinforcement_2018, fiderer_neural-network_2021}, may suggest better measurement strategies than the greedy one we applied in this paper.

From the broader perspective, we can imagine in the future to generalize the measurement setting $x$ to become an entire experimental setup, and to perform the search on a broader class of experiments.
This would be a very useful aspect of a future ``artificial scientist" that tries to explore the world and learn a model by performing experiments autonomously.


\appendix
\setcounter{section}{0}

\section{Bayesian Experimental Design}
\label{app:bayes}
In this Appendix, we describe in more detail the standard framework of Bayesian experimental design.

\subsection{The Bayesian parameter update}
Starting from some a priori distribution $P_0(\lambda)$, that reflects all our expectations on the parameters of the system (e.g. parameter range, symmetries, constraints), we can update it according to the Bayes rule \cite{papoulis_probability_2009}
\begin{equation}
    \label{app:eq:bayes}
    P_1(\lambda| x_0, y_0) = \frac{P(y_0|\lambda, x_0)P_0(\lambda)}{P_0(y_0|x_0)},
\end{equation}
where $(x_0, y_0)$ is the first measurement and its outcome and
\begin{equation}
    P_0(y_0|x_0) = \int d\lambda P(y|\lambda,x) P_0(\lambda).
\end{equation}
The distribution $P_1(\lambda)$, which depends on the performed experiment and on its outcome $(x_0,y_0)$, is called posterior distribution.
It represents the updated knowledge on $\lambda$ after taking into account the result of the experiment.
It is easy to show that in case of multiple (independent) experiments, $\left\{x_i, y_i \right\}^n$, this equation can be easily generalized to a recursive form:
\begin{align}
     & P_n \left(\lambda|\{x_i, y_i \}^n\right)= \frac{\prod^n P(y_i|\lambda, x_i) P_0(\lambda)}
    {\int d\lambda' \prod^n P(y_i|\lambda',x_i) P_0(\lambda')} =                                 \\
     & \;\;=
    \frac{P(y_n | \lambda, x_n) P_{n-1}\left(\lambda|\{x_i, y_i \}^{n-1}\right)}
    {P_{n-1}(y_n|x_n)}
\end{align}
with
\begin{equation}
    P_n(y_n|x_n) = \int d\lambda P(y_n|\lambda,x_n) P_{n-1}\left(\lambda|\{x_i, y_i \}^{n-1}\right).
\end{equation}
As one would expect, we get exactly the same updated prior if we perform many experiments in parallel and then update our knowledge or if we perform them one after the other and update our prior after each single experiment.

\subsection{The information gain query function}
We define a query function to choose which experiment to perform. This function assigns to each possible experiment $x$ its expected usefulness: the $x$ for which the query function is maximized is the one expected to be most useful to measure~\cite{settles_active_2012}.
The query function we use in this paper is the expected Kullback-Leibler divergence between the updated parameter distribution and the prior.
We start from
\begin{equation}
    KL(P_1(\lambda|y,x)||P_0(\lambda))= \int d\lambda P_1(\lambda|y,x) \log \frac{P_1(\lambda|y,x)}{P_0(\lambda)},
\end{equation}
which estimates how much the prior differs from the updated distribution as a function of the experiment $x$ and its outcome $y$.
Then, we calculate its average over the possible $y$, which gives the expected information gain when measuring at $x$ i
\begin{align}
     & IG(x) = \int dy P_0(y|x) KL(P_1(\lambda|y,x)||P_0(\lambda)) =\noindent \\
     & \;\;=
    \int dy d\lambda P_0(y|x) P_1(\lambda|y,x) \log \frac{P_1(\lambda|y,x)}{P_0(\lambda)}.
\end{align}
This equation can be rewritten as
\begin{align}
    \label{eq:app-information-gain}
     & \int dy d\lambda P_0(\lambda)P(y|\lambda,x) \log \frac{P_1(\lambda|y,x)}{P_0(\lambda)}  =\noindent \\
     & \;\;= H_0(\lambda) - H_1(\lambda|y)(x) =\noindent                                                  \\
     & \;\;= I_0(\lambda,y)(x),
\end{align}
with $ H_0(\lambda) = -\int d\lambda P_0(\lambda)\log P_0(\lambda) $ and $H_1(\lambda|y)(x) = - \int P_0(\lambda) P(y|\lambda,x)\log P_1(\lambda|y,x)$.
This can be interpreted as the mutual information between $y$ and $\lambda$, or, in other words, as the entropy reduction after measuring $(x,y)$.
The optimal $x$ to measure next, if we consider a greedy strategy, is the one that maximizes this quantity.

\section{Applications}
\label{app:examples}

\subsection{Implementation}
In our experiments, we employ a $4$-layer normalizing flow, each employing a $2$-layer MADE network (Masked Autoregressive flow for Density Estimation~\cite{papamakarios_masked_2017}) with $64$ neurons each.
After each layer of the normalizing flow, the input vector is shifted by one with a permutation layer, so that the subsequent layer applies a transformation to a different subset of the input.
We optimize the network with an Adam optimizer with learning rate $\eta=10^{-3}$.

\subsection{Coupled-cavity arrays}

We consider a cavity array with the following Hamiltonian:
\begin{equation}
    {\hat H}=\sum_j  \Omega_j {\hat a}_j^{\dagger} {\hat a}_j + \sum_j J_{j+1} ({\hat a}_j^{\dagger} {\hat a}_{j+1} + {\rm h.c.}),
\end{equation}
where the sum runs over all but the last of the cavity modes in this chain with open boundary conditions.
We consider the classical coherent-state amplitudes $a_j$ corresponding to the quantum operators ${\hat a}_j$, including drive and decay as prescribed by input-output theory~ \cite{aspelmeyer_cavity_2014}.
Given first-neighbour interactions $J_i$ and proper frequencies $\Omega_i$, we can define the matrix
\begin{equation}
    \Omega = \begin{pmatrix}
        \Omega_0 & J_1      & 0        & \ldots &          \\
        J_1      & \Omega_1 & J_2      & 0      & \ldots   \\
        0        & J_2      & \Omega_2 & J_3    & \ldots   \\
        \vdots   &          &          & \ddots &          \\
                 & \ldots   & 0        & J_N    & \Omega_N
    \end{pmatrix}.
\end{equation}
We also include some internal decay $\kappa_\text{int}$ and external decay  $\kappa_\text{ext}$, which only applies to the cavities coupled to the environment, for example the first and last.
The overall decay can be represented by a vector
\begin{equation}
    \kappa = \begin{pmatrix}
        \kappa_{\text{int}} + \kappa_{\text{ext}} \\
        \kappa_{\text{int}}                       \\
        \vdots                                    \\
        \kappa_{\text{int}}                       \\
        \kappa_{\text{int}} + \kappa_{\text{ext}}
    \end{pmatrix}.
\end{equation}
Let $a_\text{in}$ be the entering fluctuating field, which can also contain a laser drive, and $a_\text{out}$ the output field.
The behavior of the system is described by the input-output relations \cite{cohadon_quantum_2020}
\begin{equation}
    \begin{cases}
        \dot a       & = \left(- i \Omega - \frac{\kappa}{2} \right) a + \sqrt{\kappa_\text{ext}} a_\text{in} \\
        a_\text{out} & = a_\text{in} - \sqrt{\kappa_\text{ext}} a
    \end{cases}
\end{equation}
From the first equation, we can write the Green function (i.e. consider a drive $a_\text{in}\propto e^{-i\omega t}$ and assume also $a\propto e^{-i\omega t}$)
\begin{equation}
    a = \frac{\sqrt{\kappa_\text{ext}}}{-i(\omega - \Omega) + \frac{\kappa}{2}} a_\text{in}.
\end{equation}
Therefore, the final input-output relation becomes
\begin{equation}
    \label{app:eq:response}
    a_\text{out} = \left( \mathbf 1 - \frac{\sqrt{\kappa_\text{ext}}}{-i(\omega - \Omega) + \frac{\kappa}{2}}  \right) a_\text{in}.
\end{equation}

For the case of spatially constant cavity mode frequencies, the $N$ eigenfrequencies of the open-boundary array span a bandwidth $4J$.
All the frequencies in the bulk of the spectrum are doubly degenerate, and the spacing becomes smaller near the boundaries of the spectrum. This picture changes, and the degeneracies are broken, when we introduce variation in the onsite frequencies $\Omega_j$, which will be the generic situation we want to explore.


The system we discuss in the main text has couplings $J$ uniformly sampled in $[1,3]$ for each cavity, and proper frequencies sampled from a Gaussian distribution with zero mean and unit variance. The values for the specific example in~\fref{fig:measurements} and~\fref{fig:strategies} are shown in table~\ref{tab:cavities-parameters}.
We optimize the posterior neural network for $8000$ steps after each measurement, using a batch size of $1500$ samples in~\eqref{eq:information-gain0}.

\begin{table}
    \noindent\begin{minipage}[t][1\totalheight][c]{1\columnwidth}%
        \begin{center}
            \begin{tabular}{rr}
                \multicolumn{2}{c}{}                             \\
                Parameters          & Value   \tabularnewline
                \mr
                (freq.)  $\Omega_0$ & $1.040$   \tabularnewline
                $\Omega_1$          & $0.326$   \tabularnewline
                $\Omega_2$          & $0.520$   \tabularnewline
                $\Omega_3$          & $0.900$   \tabularnewline
                $\Omega_4$          & $-0.466$   \tabularnewline
                $\Omega_5$          & $0.004$   \tabularnewline
            \end{tabular}
            \hfill{}%
            \begin{tabular}{rr}
                \multicolumn{2}{c}{}                                     \\
                Known par.                   & Value   \tabularnewline
                \mr
                (coupling)     $J_1$         & $2.733$   \tabularnewline
                $J_2$                        & $2.615$   \tabularnewline
                $J_3$                        & $1.956$   \tabularnewline
                $J_4$                        & $1.568$   \tabularnewline
                $J_5$                        & $2.620$   \tabularnewline
                (int. diss.)  $\kappa_{int}$ & $0.5$   \tabularnewline
                (ext. diss.)  $\kappa_{ext}$ & $0.5$   \tabularnewline
                (msmt noise)  $\epsilon$     & $0.05$   \tabularnewline
            \end{tabular}\hfill{}%
        \end{center}
    \end{minipage}\caption{\label{tab:cavities-parameters}
        Parameters for the coupled cavity example discussed in the main text. On the left, the true frequencies we want to discover; on the right, the known parameters of the system.}
\end{table}

\subsection{Qubit chain}

Table~\ref{tab:qubits-parameters} shows the parameters employed for the qubit chain example.
We optimize the posterior neural network for at least $2500$ steps after each measurement, using a batch size of $700$ samples in~\eqref{eq:information-gain0}.
When also optimizing over $x$ (i.e. optimal measurement setting), we keep optimizing the loss until the change in $500$ steps is smaller than $0.05$.
This choice allows decreasing the training time for this more computationally-demanding system, since we train for a longer time only when required.

\begin{table}
    \noindent\begin{minipage}[t][1\totalheight][c]{1\columnwidth}%
        \begin{center}
            \begin{tabular}{rr}
                \multicolumn{2}{c}{}                            \\
                Parameters          & Value   \tabularnewline
                \mr
                (freq.)  $\omega_0$ & $1.163$   \tabularnewline
                $\omega_1$          & $1.003$   \tabularnewline
                $\omega_2$          & $1.045$   \tabularnewline
                $\omega_3$          & $0.910$   \tabularnewline
            \end{tabular}
            \hfill{}%
            \begin{tabular}{rr}
                \multicolumn{2}{c}{}                                     \\
                Known par.                     & Value   \tabularnewline
                \mr
                (coupling)     $J$             & $1.7$   \tabularnewline
                (multishot)  $n_{\text{msmt}}$ & $100$   \tabularnewline
            \end{tabular}\hfill{}%
        \end{center}
    \end{minipage}\caption{\label{tab:qubits-parameters}
        Parameters for the qubit chain example discussed in the main text. On the left, the true frequencies we want to discover; on the right, the known parameters of the system.}
\end{table}

\begin{figure}
    \centering
    \includegraphics[width=\linewidth]{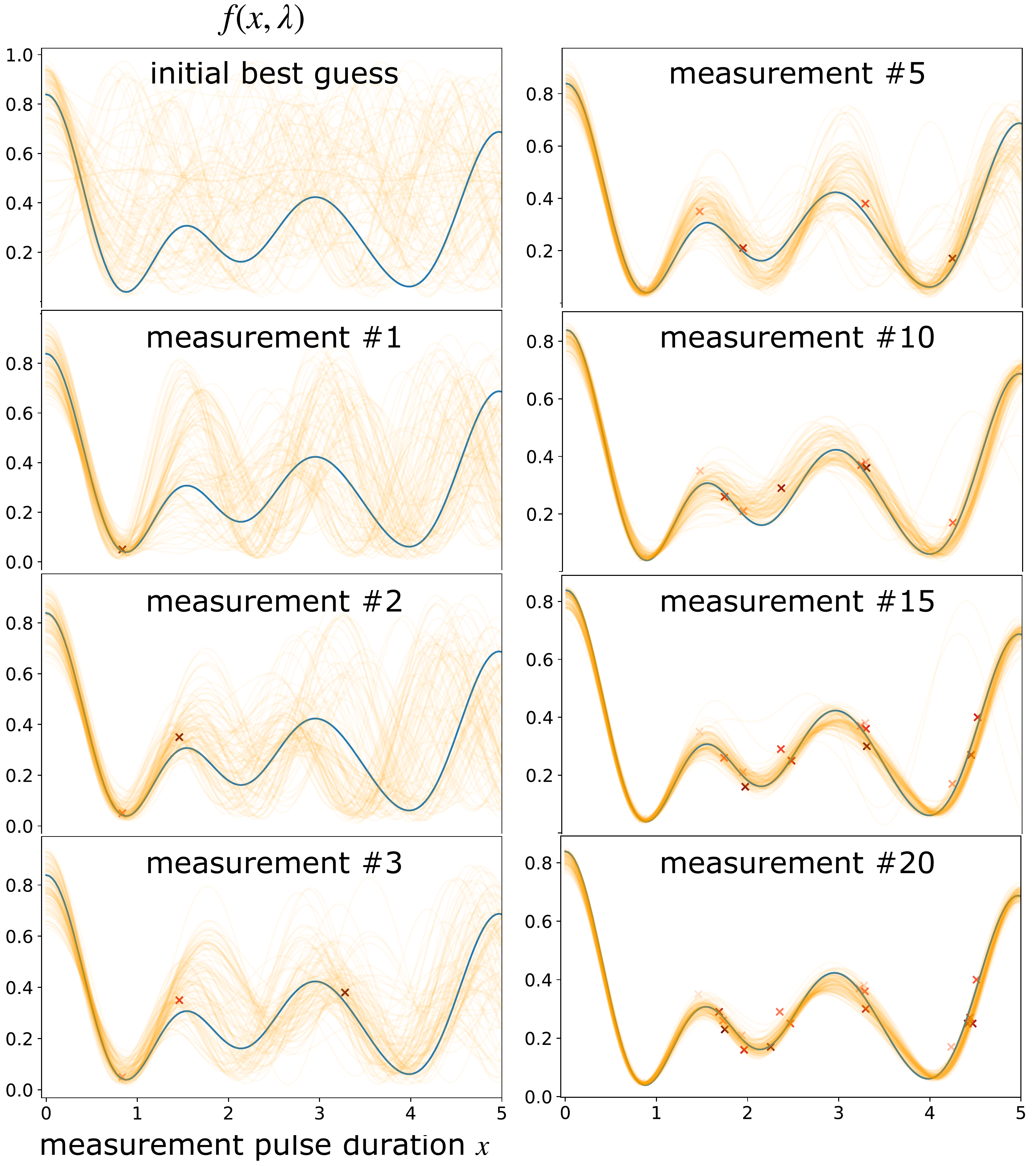}
    \caption{Prediction of the response function after a different number of steps.
        In orange, curves given by sampling parameters from the posterior; in blue, the true response function.
        The initial best guess is given by sampling from the prior distribution, before any measurement is performed.
    }
    \label{fig:app-response-evolution}
\end{figure}

It is also interesting to look at how the prediction of the response function of a qubit system as in \fref{fig:strategies-qubits} improves after the measurements suggested by an active strategy.
As shown in \fref{fig:app-response-evolution}, starting from a very uncertain prediction, the distribution of possible response functions converges to the true curve.

\medskip
\section*{References}
\bibliographystyle{iopart-num}
\bibliography{bibliography/ActiveLearning}


\end{document}